\documentstyle[epsfig,12pt]{article}
\newcommand{\be}{\begin{eqnarray}}
\newcommand{\ee}{\end{eqnarray}}

\def\lsim{\mathrel{\rlap{\lower4pt\hbox{\hskip1pt$\sim$}}
    \raise1pt\hbox{$<$}}}               
\def\gsim{\mathrel{\rlap{\lower4pt\hbox{\hskip1pt$\sim$}}
    \raise1pt\hbox{$>$}}}               

\begin{document}

\rightline{{\Large Preprint RM3-TH/02-8}}

\vspace{1cm}

\begin{center}

\LARGE{Neutron structure function and inclusive $DIS$ from $^3H$ and $^3He$ targets at large Bjorken-$x$ \footnote{Proceedings of the IX International Conference on the {\em The Structure of Baryons}, Jefferson Lab (Newport New, USA), March 3-8, 2002, World Scientific Publishing (Singapore), in press.}}

\vspace{1.5cm}

\large{M.M. Sargsian$^{(*)}$, S. Simula$^{(**)}$ and M.I. Strikman$^{(***)}$}\\

\vspace{0.5cm}

\normalsize{$^{(*)}$Department of Physics, Florida International University, Miami, FL 33199, USA\\ $^{(**)}$INFN, Sezione Roma III, Via della Vasca Navale 84, I-00146 Roma, Italy\\ $^{(***)}$Department of Physics, Pennsylvania State University, University Park, PA 16802, USA}

\end{center}

\vspace{1cm}

\begin{abstract}

\noindent A detailed study of inclusive deep inelastic scattering from mirror $A = 3$ nuclei at large values of $x_{Bjorken}$ is presented. The main purpose is to estimate the theoretical uncertainties on the extraction of $F_2^n$ from such measurements. Within the convolution approach we confirm the cancellation of nuclear effects at the level of $\approx 1\%$ for $x \lsim 0.75$ in overall agreement with previous findings. However, within models in which modifications of the bound nucleon structure functions are accounted for to describe the $EMC$ effect in nuclei, we find that the nuclear effects may be canceled at a level of $\approx 3 \%$ only, leading to an accuracy of $\approx 12 \%$ in the  extraction of $F_2^n / F_2^p$ at $x \approx 0.7 \div 0.8$. Another consequence of bound nucleon modifications is that the iteration procedure does not improve the accuracy of the extraction of $F_2^n / F_2^p$.

\end{abstract}

\newpage

\pagestyle{plain}

\section{Introduction}

\indent The investigation of deep inelastic scattering ($DIS$) of leptons off the nucleon is an important tool to get fundamental information on the structure of quark distributions in the nucleon. At large values of the Bjorken variable $x$ the ratio $d / u$, or equivalently the ratio $F_2^n / F_2^p$, is known to have a high degree of theoretical significance \cite{Isgur}. However, our present knowledge of the large-$x$ $n / p$ ratio is quite poor, mainly because its extraction from inclusive deuteron $DIS$ data is inherently model dependent. This fact has led to the suggestion of new strategies. One of them is to try to exploit the mirror symmetry of $A = 3$ nuclei; in other words, thanks to nuclear charge symmetry, one expects that the magnitude of the $EMC$ effect
  \be
      {\cal{R}}_{EMC}^A \equiv {F_2^A \over F_2^D} ~ {F_2^p + F_2^n \over Z 
      F_2^p + N F_2^n}
      \label{eq:REMC}
 \ee
is very similar in $^3He$ and $^3H$ and hence the so called super-ratio \cite{Wally}
 \be
      {\cal{SR}}_{EMC} \equiv {{\cal{R}}_{EMC}^{^3He} \over 
      {\cal{R}}_{EMC}^{^3H}} = {F_2^{^3He} \over F_2^{^3H}} ~ {2F_2^n + 
      F_2^p \over 2F_2^p + F_2^n} ~,
      \label{eq:superatio}
 \ee
should be very close to unity regardless of the size of the $EMC$ ratios \cite{Wally,Pace}. In this case the $n / p$ ratio could be extracted directly from the ratio $F_2^{^3He} / F_2^{^3H}$ without significant nuclear modifications. However, we observe that, even if nuclear charge symmetry were exact, the motion of protons and neutrons in a non isosinglet nucleus (say $^3He$) is somewhat different due to the spin-flavor dependence of the nuclear force.

\indent We have therefore performed \cite{Misak} a detailed study of inclusive $DIS$ from mirror $A = 3$ nuclei at large $x$ with the aim of estimating the theoretical uncertainties on the extraction of the $n / p$ ratio from such measurements. To this end we have considered a variety of $EMC$ models both with and without modifications of the nucleon structure function in the medium.

\section{$EMC$ models with no modification of bound nucleon}

\indent We have explored \cite{Misak} in greater details the $EMC$ model used in \cite{Wally,Pace}, which is based on the Virtual Nucleon Convolution ($VNC$) approach with no modification of the nucleon structure function in the medium. 

\indent We have analyzed the effects of: ~ i) charge symmetry breaking terms in the nucleon-nucleon ($NN$) interaction; ~ ii) finite $Q^2$ effects in the impulse approximation; ~ iii) the role of different prescriptions for the nucleon Spectral Function  normalization providing baryon number conservation; and ~ iv) the role of different parton distribution function sets. Additionally we have also compared the predictions of the $VNC$ model with the ones obtained within the light-cone ($LC$) formalism.

\indent Within the $VNC$ model, in which no modification of the bound nucleon is considered, the deviation of the super-ratio (\ref{eq:superatio}) from unity is found to stay within $1 \%$ only for $x \lsim 0.75$ in agreement with \cite{Wally,Pace} in which the effects of (i)-(iv) have not been considered.

\section{$EMC$ models with $F_2^N$ modifications in the medium}

\indent However the above estimations are not complete, since $VNC$ and $LC$ models are just two of the many models of the $EMC$ effect and, moreover, they underestimate significantly the $EMC$  data at large $x$ for a variety of nuclei. In particular, the convolution approach is not able to reproduce the minimum of the $EMC$ ratio around $x \approx 0.7$ as well as the subsequent sharp rise at larger $x$ (see \cite{Misak}). {\em Thus in order to draw final conclusions about the size of the deviation of the super-ratio ${\cal{SR}}_{EMC}$ from unity one should investigate effects beyond those predicted by the convolution approach.}

\indent Furthermore, it is very important to asses any isospin dependence of the $EMC$ effect in order to extract $F_2^n$ from $^3He$ and $^3H$ data in a reliable way. An isospin dependence of the $EMC$ effect is naturally expected from the differences in the relative motion of $pn$ and $nn$ ($pp$) pairs in $^3H$ ($^3He$). Since the interaction of a $pn$ pair is more attractive than the one of a $nn$ pair, the overlapping probability to find a $NN$ pair with $r_{NN} \leq 1 ~ fm$, may be $\sim 40\%$ larger for a $pn$ pair than for a $nn$ pair (see \cite{Pieper}). This is a very important isospin effect in mirror $A = 3$ nuclei, which can lead to deviations of the super-ratio (\ref{eq:superatio}) from unity depending on the size of the $EMC$ effect itself. {\em It is worth noting that the nuclear charge symmetry will not limit such deviations.}

\indent Hence we have carried out a detailed analysis of the super-ratio within a broad range of models of the $EMC$ effect, which take into account possible modifications of the bound nucleons in nuclei, like: ~ i) a change in the quark confinement size (including swelling); ~ ii) the possible presence of clusters of six quarks; and ~ iii) the suppression of point-like configurations due to color screening. Our main result \cite{Misak} is that one cannot exclude the possibility that the cancellation of the nuclear effects in the super-ratio may occur only at a level of $\approx 3 \%$, resulting in a significant uncertainty (up to $\approx 12 \%$ for $x \approx 0.7 \div 0.8$) in the extraction of the free $n / p$ ratio from the ratio of the measurements of the $^3He$ and $^3H$ $DIS$ structure functions (see Fig. 1). 

\indent In \cite{Pace} it was suggested that, once the ratio $F_2^{^3He} / F_2^{^3H}$ is measured, one can employ an iterative procedure to extract the $n / p$ ratio which can almost eliminate the effects of the dependence of the super-ratio ${\cal{SR}}_{EMC}$ on the large-$x$ behavior of the specific structure function input. Namely, after extracting the $n / p$ ratio assuming a particular calculation of ${\cal{SR}}_{EMC}$, one can use the extracted $F_2^n$ to get a new estimate of ${\cal{SR}}_{EMC}$, which can then be employed for a further extraction of the $n / p$ ratio. Such a procedure can be iterated until convergence is achieved and self-consistent solutions for the extracted $F_2^n / F_2^p$ and the super-ratio ${\cal{SR}}_{EMC}$ are obtained. In \cite{Pace} a good convergence was achieved for $x$ up to $\simeq 0.8$. However this result depends on the assumed validity of the convolution approach at large $x$. Within the $EMC$ models considered above we have found \cite{Misak} that: ~ i) the consistency between the $n / p$ ratio used as an input and the extracted one is not guaranteed; ~ ii) the iteration diverges already at values of $x$ ($\simeq 0.7$) smaller than the ones obtained in \cite{Pace}; moreover, below $x \simeq  0.7$ the iteration procedure converges to a value of  the $n / p$ ratio which differs from the input one exactly by the amount of the $EMC$ effect implemented in the calculation of $F_2^{^3He} / F_2^{^3H}$.

\begin{figure}[htb]

\centerline{\epsfxsize=16cm \epsfig{file=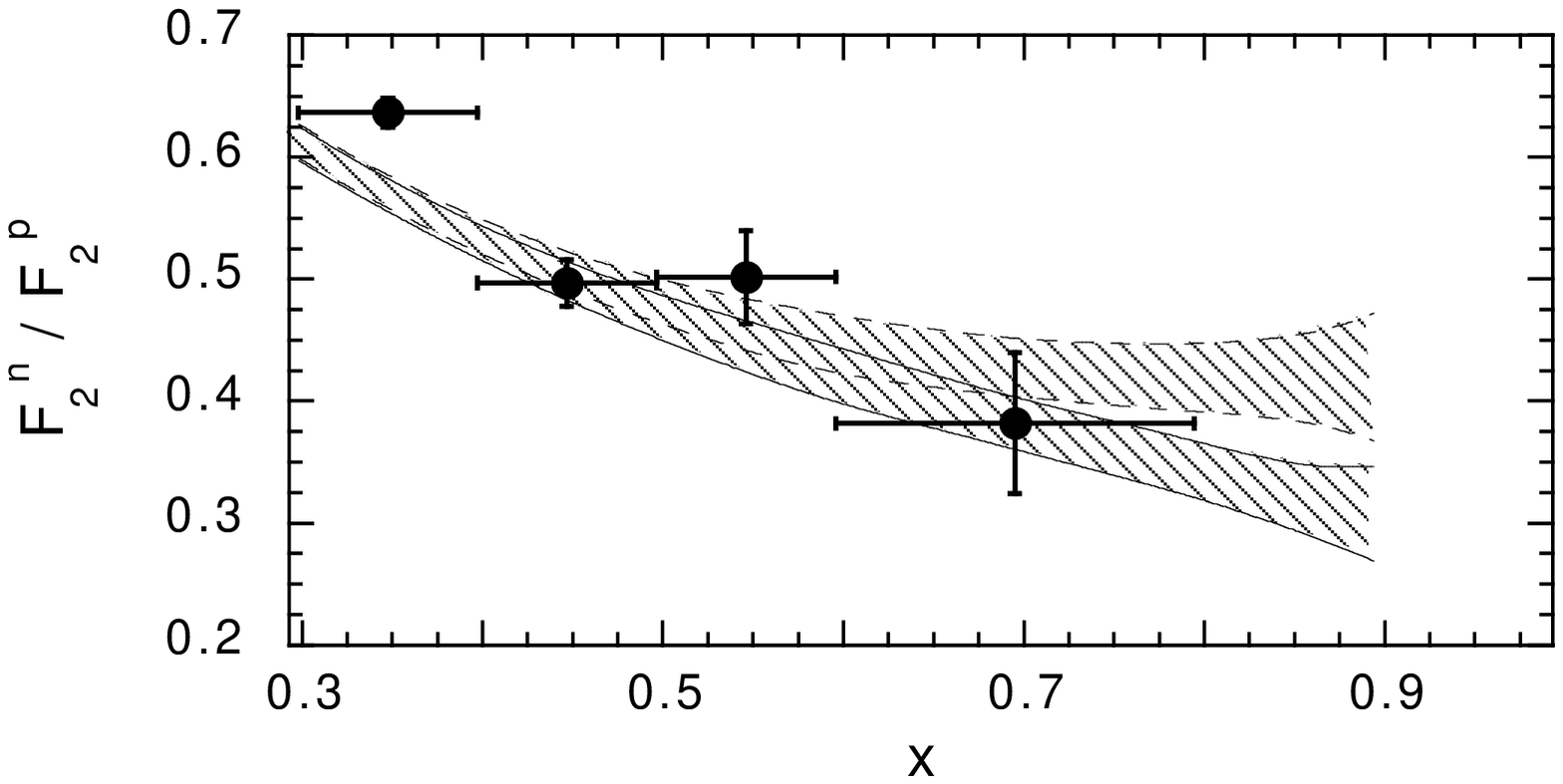}}

{\small {\bf Figure 1.} The expected accuracy for the extraction of $F_2^n / F_2^p$ vs. $x$ at $Q^2 = 10 ~ (GeV/c)^2$. The lower and upper shaded areas correspond to the $CTEQ$ and modified $CTEQ$ parameterizations as described in \protect\cite{Misak}. Full dots are $NMC$ data \protect\cite{Arneodo}.}

\end{figure}

\indent Thus we conclude that in spite of the presented limitations the measurements with mirror $A = 3$ nuclei will significantly improve our knowledge of the $n / p$ ratio at large $x$. However, it will be very important to complement them with the measurements of semi-inclusive processes off the deuteron, in which the momentum of the struck nucleon is tagged by detecting the recoiling one. Imposing the kinematical conditions that the detected momentum is low ($\lsim 150 ~ MeV/c$), which means that the nucleons in the deuteron are initially far apart \cite{SIM96,Strikman}, it is possible to minimize significantly the nuclear effects. Furthermore, all the unwanted nuclear effects can be isolated by using the same reaction for the extraction of the proton structure function by detecting slow recoiling neutrons and comparing the results with existing hydrogen data, as well as by performing tighter cuts on the momentum of the spectator proton and then extrapolating to the neutron pole.

\end{document}